\documentclass[aps,pra,superscriptaddress,twocolumn,longbibliography]{revtex4-1}
\usepackage{units}
\usepackage{amsmath}
\usepackage{amsthm}
\usepackage{amssymb}
\usepackage{graphicx}
\usepackage{color}
\usepackage{bbold}

\usepackage[colorlinks=true, urlcolor=blue, citecolor=blue,linkcolor=blue,citebordercolor={1 0 0},linkbordercolor={0 0 1}]{hyperref}

\graphicspath{{./images/},{./imagesAppendix/}}

\theoremstyle{plain}
\newtheorem{thm}{\protect\theoremname}
\theoremstyle{plain}

\ifx\proof\undefined
\newenvironment{proof}[1][\protect\proofname]{\par
	\normalfont\topsep6\p@\@plus6\p@\relax
	\trivlist
	\itemindent\parindent
	\item[\hskip\labelsep\scshape #1]\ignorespaces
}{%
	\endtrivlist\@endpefalse
}
\providecommand{\proofname}{Proof}
\fi
\theoremstyle{plain}

\theoremstyle{remark}

\newcommand{\bra}[1]{\langle #1|}
\newcommand{\ket}[1]{|#1 \rangle}
\makeatother
\newcommand{\braket}[2]{\langle #1 \vert #2 \rangle}
\newcommand{\abs}[1]{\left|#1\right|}
\newcommand{\idg}[1]{{\bfseries #1)}}

\newcommand{\revB}[1]{#1}
\newcommand{\revC}[1]{{#1}}
\newcommand{\Tr}{\mathrm{Tr}}

\newcommand\numberthis{\addtocounter{equation}{1}\tag{\theequation}}
\providecommand{\factname}{Fact}
\providecommand{\theoremname}{Theorem}
\providecommand{\claimname}{Claim}
\providecommand{\lemmaname}{Lemma}
\providecommand{\definitionname}{Definition}

\definecolor{KB}{rgb}{0.4,0.3,0.9}

\definecolor{THc}{rgb}{0.9,0.3,0.2}

\theoremstyle{definition}
\newtheorem{defn}[thm]{\protect\definitionname}

\newcommand{\subfigimg}[3][,]{%
	\setbox1=\hbox{\includegraphics[#1]{#3}}
	\leavevmode\rlap{\usebox1}
	\rlap{\hspace*{2pt}\raisebox{\dimexpr\ht1-0.5\baselineskip}{{\bfseries \large\textsf{#2}}}}
	\phantom{\usebox1}
}

\begin{document}
\title{Generalized Quantum Assisted Simulator}
\author{Tobias Haug}
\email{thaug@ic.ac.uk}
\affiliation{Centre for Quantum Technologies, National University of Singapore 117543, Singapore}
\author{Kishor Bharti}
\email{kishor.bharti1@gmail.com}
\affiliation{Centre for Quantum Technologies, National University of Singapore 117543, Singapore}
\begin{abstract}
We provide a noisy intermediate-scale quantum framework for simulating the dynamics of open quantum systems,  generalized time evolution, non-linear differential equations and Gibbs state preparation.
Our algorithm does not require any classical-quantum feedback loop, bypass the barren plateau problem and does not necessitate any complicated measurements such as the Hadamard test. We introduce the notion of the hybrid density matrix, which allows us to disentangle the different steps of our algorithm and delegate classically demanding tasks to the quantum computer. Our algorithm proceeds in three disjoint steps. First, we select the ansatz, followed by measuring overlap matrices on a quantum computer. The final step involves classical post-processing data from the second step. 
Our algorithm has potential applications in solving the Navier-Stokes equation, plasma hydrodynamics, quantum Boltzmann training, quantum signal processing and linear systems.
Our entire framework is compatible with current experiments and can be implemented immediately.
\end{abstract}
\maketitle

\section{Introduction}
The quest for quantum advantage for practical use-cases in the noisy
intermediate-scale quantum (NISQ)  era~\cite{preskill2018quantum,deutsch2020harnessing} has spurred the development
of algorithms, which can be executed on shallow quantum circuits and
do not necessitate error correction. Despite the hope rendered by
the recent Google quantum supremacy experiment~\cite{arute2019quantum} at the hardware frontier,
it remains to devise algorithms which can harness the power of the
NISQ devices for problems of practical relevance.

The task of estimating the ground state and ground state energy of
a Hamiltonian is one such model problem, for which one could expect
to conceive algorithms for quantum advantage. Another canonical
problem
is the broader challenge of simulating the quantum dynamics. 
In fact, the birth of the field of quantum computation can be attributed to
Feynman's dream of simulating the quantum dynamics~\cite{feynman1982simulating}. While the Hamiltonian
ground state problem has applications in combinatorial optimization,
solid-state physics and quantum chemistry, quantum simulation offers
the possibility to explore topics such as high-temperature superconductivity
and drug design. Interestingly, both Hamiltonian ground state problem
and quantum simulation can be tackled via variational principles,
based on static and dynamical methods respectively.

The leading canonical NISQ era algorithm for approximating the ground state and ground state energy of a Hamiltonian is variational quantum eigensolver (VQE)~\cite{peruzzo2014variational,mcclean2016theory,kandala2017hardware,farhi2014quantum,farhi2016quantum,harrow2017quantum, farhi2016quantum,mcardle2020quantum,endo2020hybrid}. The aforementioned algorithm is based on the Rayleigh-Ritz variational principle and employs a classical-quantum feedback loop to update the parameters of the corresponding parametric quantum circuit (PQC). The classical optimization program corresponding to VQE is highly non-convex and in general uncharacterized~\cite{bittel2021training}. The classical-quantum feedback loop further impedes the possibility to utilize the quantum computer, until the classical computer has calculated its output. In general the ansatz is either not compatible with the existing hardware capabilities or chosen in a heuristic fashion. The absence of a mathematically rigorous structure renders any systematic investigation challenging. Moreover, the recent results revealing the existence of the barren plateau as the hardware noise,  number of qubits or amount of entanglement increase, has led to genuine concerns about the fate of VQE~\cite{mcclean2018barren,huang2019near,sharma2020trainability,cerezo2020cost,wang2020noise,marrero2020entanglement}. Even gradient-free optimization techniques fail to evade the fatality of the barren plateaus~\cite{Arrasmith2020Effect}.

For simulating the dynamics of closed quantum systems, the leading canonical NISQ era algorithm is the variational quantum simulation (VQS) algorithm~\cite{li2017efficient,mcardle2019variational,yuan2019theory}.
The VQS algorithm utilizes a hybrid classical-quantum feedback loop to update the parameters of a PQC using dynamical variational principles. 
The aforementioned algorithm as well as its VQE based variant, i.e. the subspace variational quantum simulator (SVQS)~\cite{heya2019subspace}, share resemblances and hence many of the problems of VQE such as the barren plateau problem.  The VQS algorithm furthermore requires complicated measurements, does not provide a systematic strategy to choose the ansatz and mandates the adjustable parameters to be real-valued~\cite{yuan2019theory}.

Recently algorithms as alternative beyond VQE and VQS were proposed in the literature~\cite{bharti2020quantum,bharti2020iterative,bharti2020quantum2}. The quantum assisted eigensolver (QAE) and iterative quantum assisted eigensolver (IQAE) demonstrate a systematic structure in their algorithm ~\cite{bharti2020quantum,bharti2020iterative}. 
Their classical optimization program is well-characterized quadratically constrained quadratic program with a single equality constraint. In particular, the IQAE algorithm offers an organized path to construct the ansatz, circumvents the barren plateau problem, does not mandate any quantum-classical feedback loop and can be efficiently executed on the current quantum hardware without the need of complicated measurements. 
To tackle the challenges encountered by VQS, the quantum assisted simulator (QAS) can simulate dynamics of quantum systems while boasting the same advantages as IQAE~\cite{bharti2020quantum2}.

The task of simulating the dynamics for open quantum systems is relatively more demanding than for closed quantum systems. Here,  challenging tasks related to dynamics are generalized time evolution with a non-Hermitian or nonlinear Hamiltonian as well as Gibbs state preparation.  
In the literature, various algorithms have been proposed to solve the aforementioned tasks~\cite{endo2020variational,kyriienko2020solving,lubasch2020variational,gaitan2020finding,lloyd2020quantum,chowdhury2020variational,hu2020quantum,yoshioka2020variational,liu2020solving}. However, the existing algorithms are either not compatible with current hardware capabilities or share the troubles faced by VQS.

In this work, we provide the generalized quantum assisted simulator (GQAS) to simulate open system dynamics, generalized time evolution, nonlinear differential equations and Gibbs state preparation. The GQAS algorithm furnish an antidote to the hardships faced by the current NISQ alternatives. In particular, the GQAS does not mandate any classical-quantum feedback loop, circumvents the barren plateau problem and does not require any complicated measurements. The whole framework is compatible with existing hardware capabilities. 

\begin{figure}[htbp]
	\centering
	\subfigimg[width=0.49\textwidth]{}{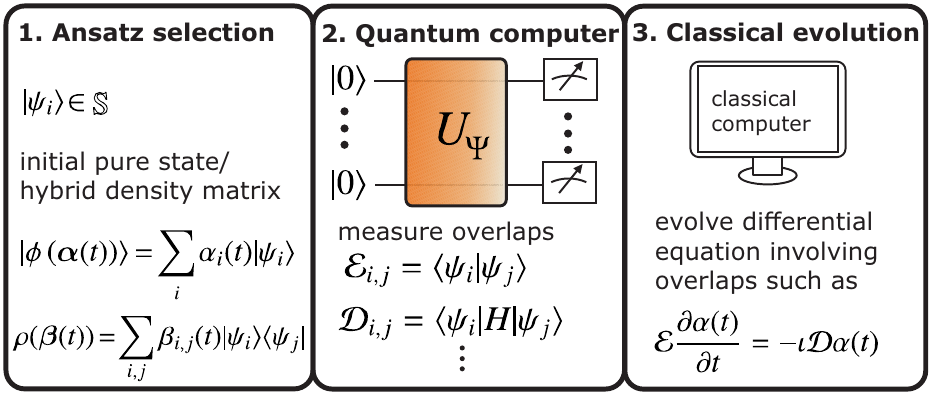}
	\caption{Concept of the general quantum assisted algorithm (GQAS). It consists of three steps. The first step selects the ansatz as a linear combination of states $\ket{\psi_i}\in\mathbb{S}$ from a set $\mathbb{S}$. The ansatz can be either a pure state or a mixed state $\rho$. 
	We introduce the concept of hybrid mixed state $\rho=\sum_{i,j}\beta_{i,j}(t)\ket{\psi_i}\bra{\psi_j}$. Only the coefficients $\beta_{i,j}(t)$ which are stored on a classical computer are varied in time, thus avoiding any classical-quantum feedback loop.
	The second step computes the overlap matrices on a quantum computer, which can be measured efficiently using Pauli strings~\cite{bharti2020quantum2}.  Finally, the differential equation to be computed is solved on a classical computer. 
	}
	\label{QASComic}
\end{figure}

\revC{We introduce our algorithm in Sec.~\ref{sec:alg}. Then, we show four different applications of our algorithm. First, we solve open system dynamics in Sec.\ref{sec:open}, then generalized time evolution in Sec.\ref{sec:gtime}, followed by nonlinear differential equations in Sec.\ref{sec:nld} and Gibbs state preparation in Sec.\ref{sec:gibbs}. Finally, we discuss the results in Sec.\ref{sec:discussion}.}

\section{Algorithm}\label{sec:alg}

\revC{The execution of GQAS algorithm involves in general three steps (see Fig.\ref{QASComic} for pictorial synopsis), which we show in the following. The specific details of GQAS vary depending on the application, which we defer to the application part in Secs.~\ref{sec:open}-\ref{sec:gibbs}. Our algorithm proceeds as follows}
\begin{enumerate}
\item The ansatz is selected as a linear combination of quantum states
\item Calculation of the overlap matrices on a quantum computer
\item Solving the dynamical evolution equation on a classical computer
\end{enumerate}
\revC{First, we choose a set of quantum states $\mathbb{S}=\{\ket{\psi_j} \}_j$. The states should be chosen such that they span the space of the problem, and can be selected adapted to the application as shown in Sec.\ref{sec:open}. The ansatz is either a linear combination of states $\ket{\phi}=\sum_{i=1}^M\alpha_i\ket{\psi_i}$ with classical combination coefficients $\alpha_i\in\mathbb{C}$ (Sec.\ref{sec:gtime},\ref{sec:nld}) or a hybrid density matrix with a coefficient matrix $\beta_{i,j}\in\mathbb{C}$ as introduced in Eq.\ref{def:Virtual_density_matrix} (Sec.\ref{sec:open},\ref{sec:gibbs}).
Note that the ansatz states are fixed and only classical combination coefficients are updated later on. } 

\revC{For step $2$, the quantum computer measures overlap matrices of the form $\bra{\psi_i}O\ket{\psi_j}$ with some observable $O$. As we will show in Sec.~\ref{sec:open},  step $2$ can be performed efficiently on a quantum computer, without the requirement of any complicated measurements such as the Hadamard test. When the ansatz states $\ket{\psi_j}=P_j\ket{\psi}$ are generated by applying a set of Pauli strings $\{P_j\}_{j=1}^M$ to a reference state $\ket{\psi}$, the overlaps can be calculated as simple measurements of Pauli strings of the form $\bra{\psi}P\ket{\psi}$ with some Pauli string $P$. This is because products of Pauli strings are again Pauli strings up to a pre-factor $\pm 1, \pm \iota$, which can be calculated trivially.
}

\revC{As last step, we perform evolve the combination coefficients using the measured overlaps. Note that the equations depend on the particular application and will be discussed in detail in the following sections.  
}

\subsection*{Hybrid Density Matrix}
Here, we introduce the concept of the hybrid density matrix which will be used for applications in Sec.\ref{sec:open},\ref{sec:gibbs}. The ansatz is constructed from a set of fixed quantum states, which we define as follows
\begin{defn}  \label{def:Virtual_density_matrix}
\textbf{Hybrid density matrix:} Given a Hilbert space $\mathcal{H}$ and a set of $M$ quantum states $\mathbb{S}=\{\ket{\psi_j} \in \mathcal{H} \}_j$, a hybrid density matrix $\rho$ is given by
\begin{equation}\label{eq:open_ansatz}
\rho=\sum_{\left(\ket{\psi_i},\ket{\psi_j}\right) \in \mathbb{S} \times \mathbb{S}}\beta_{i,j}\ket{\psi_i}\bra{\psi_j}\,
\end{equation}
for  $\beta_{i,j} \in \mathbb{C}$. The coefficients ($\{\beta_{i,j}\}_{i,j}$) are stored on a classical device and the quantum states correspond to some quantum system. A hybrid density matrix represents a valid density matrix if $\text{Tr}(\rho)=1$ and $\rho \succeq 0$.
\end{defn}
The normalization condition is fulfilled when
\begin{equation}
\text{Tr}(\rho)=\text{Tr}(\boldsymbol{\beta} \mathcal{E})=1\,,
\end{equation}
where the coefficient matrix $\boldsymbol{\beta}$ is a positive semidefinite matrix and we define the overlap matrix via
\begin{equation}
\mathcal{E}_{i,j}=\langle\psi_{i}\vert\psi_{j}\rangle. \label{eq:open_E}
\end{equation}
\revB{The positive semidefinite condition of $\rho$ is automatically fulfilled when $\boldsymbol{\beta}$ is a positive semidefinite matrix as we have \begin{equation}
\bra{x}\rho\ket{x}=\sum_{i,j}\braket{x}{\psi_i}\beta_{i,j}\braket{\psi_j}{x}=c_x^\dagger \boldsymbol{\beta} c_x\ge0  \,\,\,\,\,\,\forall\, \ket{x}\,,
\end{equation}
where $c_x$ is a vector with $c_x^j=\braket{\psi_j}{x}$.}
We note that the purity of the hybrid density matrix is given by $\text{Tr}(\rho^2)=\text{Tr}(\mathcal{E}\boldsymbol{\beta}\mathcal{E}\boldsymbol{\beta})$.
The notion of hybrid density matrix helps us remove the quantum-classical feedback loop and thus renders the different steps of our GQAS algorithm disjoint. \revB{Here, the hybrid density matrix can be classically updated by tuning the classical coefficients $\beta_{i,j}$  without requiring a change to the quantum states $\ket{\psi_i} \in \mathcal{H}$ on the quantum computer.} \revB{We note that expressing the ansatz state via a classical combination of quantum states is a powerful concept that has been used in a number of papers~\cite{ollitrault2020quantum,bharti2020iterative,seki2021quantum,huggins2020non,stair2020multireference,parrish2019quantum2,mcclean2017hybrid,yuan2021quantum,endo2021hybrid} for calculating ground and excited states. A representation of density matrices with variational quantum circuits has been proposed in reference~\cite{endo2021hybrid,yoshioka2020variational}. }

In the following sections, we proceed to discuss GQAS algorithm for simulating the dynamics of open systems, generalized time evolution, solving nonlinear differential equations and Gibbs state preparation.

\section{Open System Dynamics}\label{sec:open}
A system interacting with a bath within the Born-Markov approximation can be described with the Lindblad master equation~\cite{breuer2002theory}
\begin{equation}\label{eq:Lindblad}
\frac{d}{dt}\rho=-\iota[H,\rho]+\sum_{n=1}^f \gamma_n(L_n\rho L_n^\dagger -\frac{1}{2}L_n^\dagger L_n\rho-\frac{1}{2}\rho L_n^\dagger L_n )\,,
\end{equation}
where $\rho$ is the density matrix of the system, $H$ describes the Hamiltonian of the system and $f$ operators $L_k$ encode the action of the bath on the system and $\gamma_n \geq 0$. This type of equation has been employed to describe a wide range of systems interacting with the environment. This equation is valid as long as the interaction between system-bath is weak, and the correlation between system and bath decay fast.

To evolve the open system problem, we now introduce the following notations
\begin{subequations}
 \label{eq:open_all}
\begin{align}
\mathcal{D}_{i,j}=&\langle\psi_{i}\vert H\vert\psi_{j}\rangle. \label{eq:open_D}\\
\mathcal{R}_{i,j}^n=&\langle\psi_{i}\vert L_n\vert\psi_{j}\rangle. \label{eq:open_R}\\
\mathcal{F}_{i,j}^n=&\langle\psi_{i}\vert L_n^\dagger L_n\vert\psi_{j}\rangle. \label{eq:open_F}
\end{align}
\end{subequations}
Using above overlap matrices and Eq.\eqref{eq:open_ansatz}, we can now use the  Dirac and Frenkel variational principle~\cite{yuan2019theory} 
\begin{equation}
\text{Tr}(\delta \rho (\frac{d}{dt}\rho-\mathcal{L}(\rho)))=0\,,
\end{equation}
where $\mathcal{L}(\rho)$ is the right hand side of Eq.\eqref{eq:Lindblad}. We then find the following differential equation for the time dependent parameters $\boldsymbol{\beta}(t)$,
\begin{align*}
\mathcal{E}\frac{\text{d}}{\text{d}t}&\boldsymbol{\beta}(t)\mathcal{E}=-\iota(\mathcal{D}\boldsymbol{\beta}(t)\mathcal{E}-\mathcal{E}\boldsymbol{\beta}(t)\mathcal{D})+\\
&\sum_{n=1}^f \gamma_n (\mathcal{R}_n\boldsymbol{\beta}(t)\mathcal{R}_n^\dagger-\frac{1}{2}\mathcal{F}_n\boldsymbol{\beta}(t)\mathcal{E}-\frac{1}{2}\mathcal{E}\boldsymbol{\beta}(t)\mathcal{F}_n).\numberthis\label{eq:Lindblad_E}
\end{align*}
Assuming that $H=\sum_i a_i U_i$ and $L_k=\sum_i b_i V_i$ is a linear combination of $N$-qubit unitaries $U_i$ and $V_i$, where each unitary acts non-trivially on at most $\mathcal{O}\left(poly\left(logN\right)\right)$ qubits, then the overlap matrices can be measured efficiently using methods of \cite{mitarai2019methodology}. For $U_i$, $V_i$ being Pauli strings, the overlaps can be easily calculated as measurements of Pauli strings and one can relax the aforementioned  $\mathcal{O}\left(poly\left(logN\right)\right)$ constraint. 

Note that for an ansatz that does not cover the full Hilbert space the evolution with the Lindblad terms may not preserve the trace of the ansatz $\text{Tr}(\mathcal{E}\boldsymbol{\beta})$. To accommodate for this, we numerically normalize $\boldsymbol{\beta}(t)$ during the classical integration after every time step.

We show an example dissipative problem in Fig.\ref{ExamleDissp} which first has been discussed in~\cite{endo2020variational}. 
The Hamiltonian is an Ising model in a ladder configuration
\begin{equation}\label{eq:ising_ladder}
H_\text{L}=\sum_{\langle i,j \rangle} J\sigma^z_i\sigma^z_j+\sum_i h\sigma_i^x
\end{equation}, where $\langle i,j \rangle$ denotes the set of nearest-neighbor couplings, $\sigma^z_i$ is the $z$ Pauli operator acting on the $i$-th qubit. The qubits are arranged in a ladder configuration. Dissipation acts on the system in form of a spontaneous creation of excitations realized by $L_i=\sqrt{\gamma}\sigma^+$, where $\gamma$ is the creation rate and $\sigma^+=\ket{1}\bra{0}_i$ is the raising operator acting on qubit $i$. 

As first step, we choose a problem-aware ansatz that captures the evolving subspace by using the Hamiltonian. 
We generate a set of $M$ basis states $\ket{\psi_i}$ with the $K$-moment expansion~\cite{bharti2020quantum2}. \revB{This method of construction is inspired by the Krylov subspace expansion, which uses an expansion in terms of higher orders of the Hamiltonian $H$ to represent the solution space~ $\text{span}(\psi,H\psi,H^2\psi,\dots,H^K\psi)$~\cite{saad1992analysis}. The $K$-moment expansion is an adaption of this method which is suitable for NISQ devices.} Here, the basis states are taken from the cumulative $K$-moment states~\cite{bharti2020quantum2}
\begin{equation}
\mathbb{CS}_{K}=\{\vert \psi \rangle\} \cup \left\{ U_{i_1}\vert\psi\rangle\right\} _{i_1=1}^{r} \cup \dots \cup \left\{ U_{i_K}\dots U_{i_1}\vert\psi\rangle\right\} _{i_1,\dots,i_K}^{r}\, ,\label{eq:kmoment}
\end{equation}
where $\ket{\psi}$ is some efficiently preparable reference state and $U_i$ are taken from the set of $r$ Pauli strings that make up Eq.\eqref{eq:ising_ladder} with $H=\sum_{i=1}^r a_i U_i$.

As second step, we calculate the overlap matrices given in Eq.\eqref{eq:open_all}. This step is performed on a quantum computer, which we simulate here numerically. Our choice of $U_i$ as Pauli strings allows us to determine all the overlaps as simple measurements of Pauli strings, which can be efficiently done on the current quantum hardware.
We choose the initial state for the evolution to be $\ket{\phi}=\ket{0}^{\otimes N}$. 
We have to find the parameters $\beta_{i,j}(t=0)$ with $\ket{\phi}\bra{\phi}=\sum_{i,j}\beta_{i,j}(0)\ket{\psi_i}\bra{\psi_j}$ that approximate this initial state. 
We find the initial state using IQAE~\cite{bharti2020iterative}, where we use the same overlaps already calculated for GQAS. We define the Hamiltonian $H_\text{ini}=-\sum_{i=1}^N\sigma^z_i$, with the ground state of $H_\text{ini}$ being the initial state $\ket{\phi}$. IQAE can now be applied to find the ground state by minimizing $\boldsymbol{\alpha}$ in respect to $\boldsymbol{\alpha^\dag} \mathcal{G}\boldsymbol{\alpha}$ under the condition $\boldsymbol{\alpha^\dag} \mathcal{E}\boldsymbol{\alpha}=1$, where $\mathcal{G}_{i,j}=\langle\psi_{i}\vert H_\text{ini}\vert\psi_{j}\rangle$. With the minimized $\boldsymbol{\alpha}_\text{min}$, we then construct $\beta_{i,j}(t=0)=\alpha_i\alpha_j^*$. 

As third and final step, we use Eq.\eqref{eq:Lindblad_E} to solve the dynamics classically using the measured overlaps.

\begin{figure}[htbp]
	\centering
	\subfigimg[width=0.26\textwidth]{}{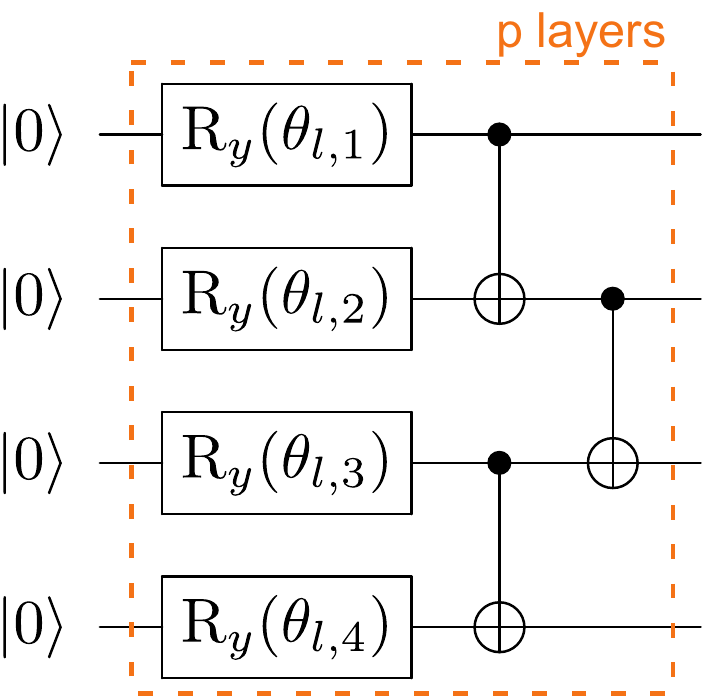}
	\caption{Circuit of $N$ qubits to generate reference state $\ket{\psi}$. This state is used to generate ansatz states $\{\ket{\psi_j}=P_j\ket{\psi}\}_{j=1}^M$ by applying different Pauli strings $P_j$. $\ket{\psi}$ consists of $p$ layers of $N$ single qubit rotations around the $y$ axis with parameters $\theta_{l,i}$, followed by CNOT gates arranged in a nearest-neighbor chain topology. The parameters $\theta_{l,i}$ are chosen at random and then kept fixed for the entirety of the GQAS algorithm.
	}
	\label{AnsatzCircuit}
\end{figure}

\begin{figure*}[htbp]
	\centering
	\subfigimg[width=0.3\textwidth]{a}{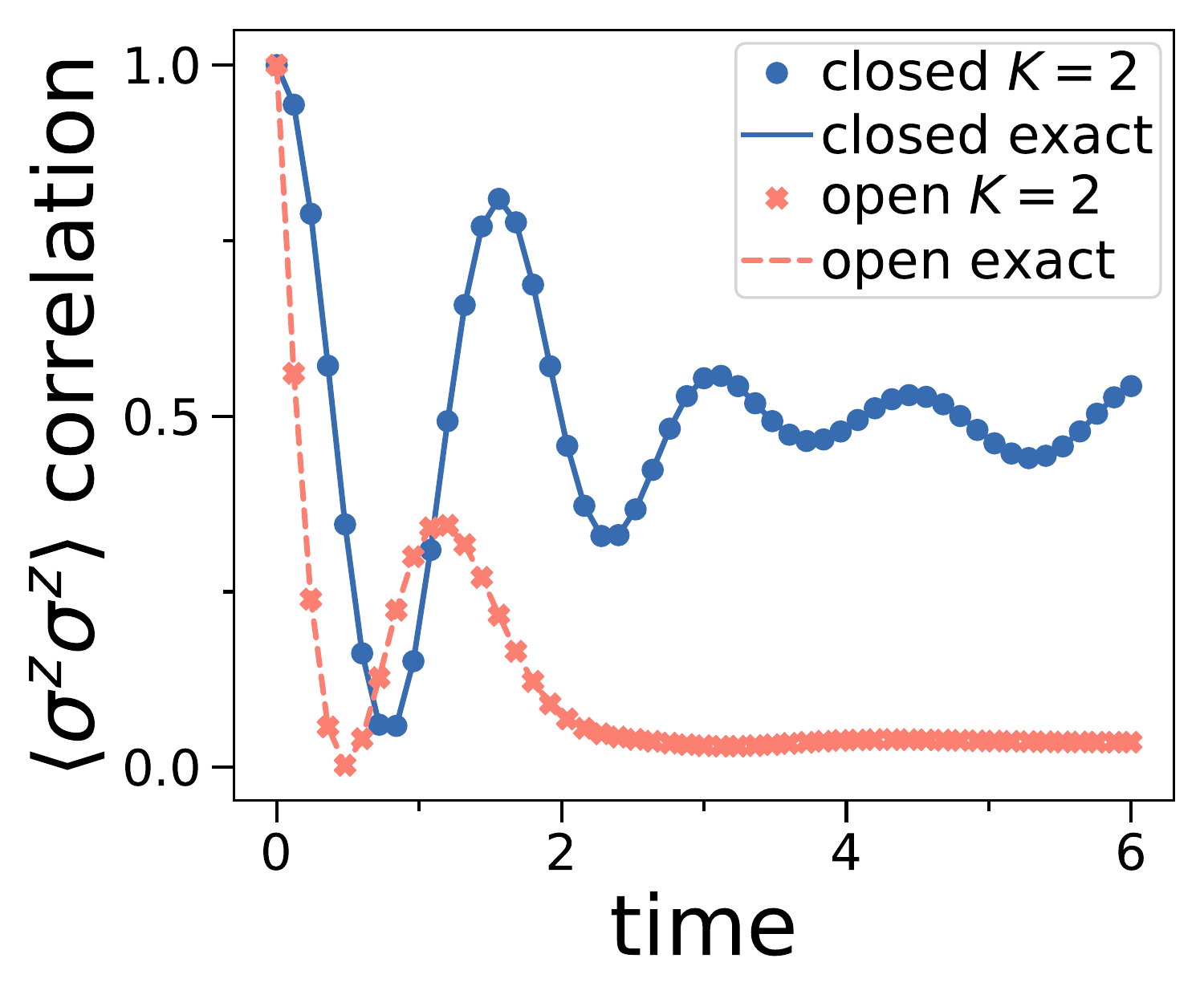}
	\subfigimg[width=0.68\textwidth]{}{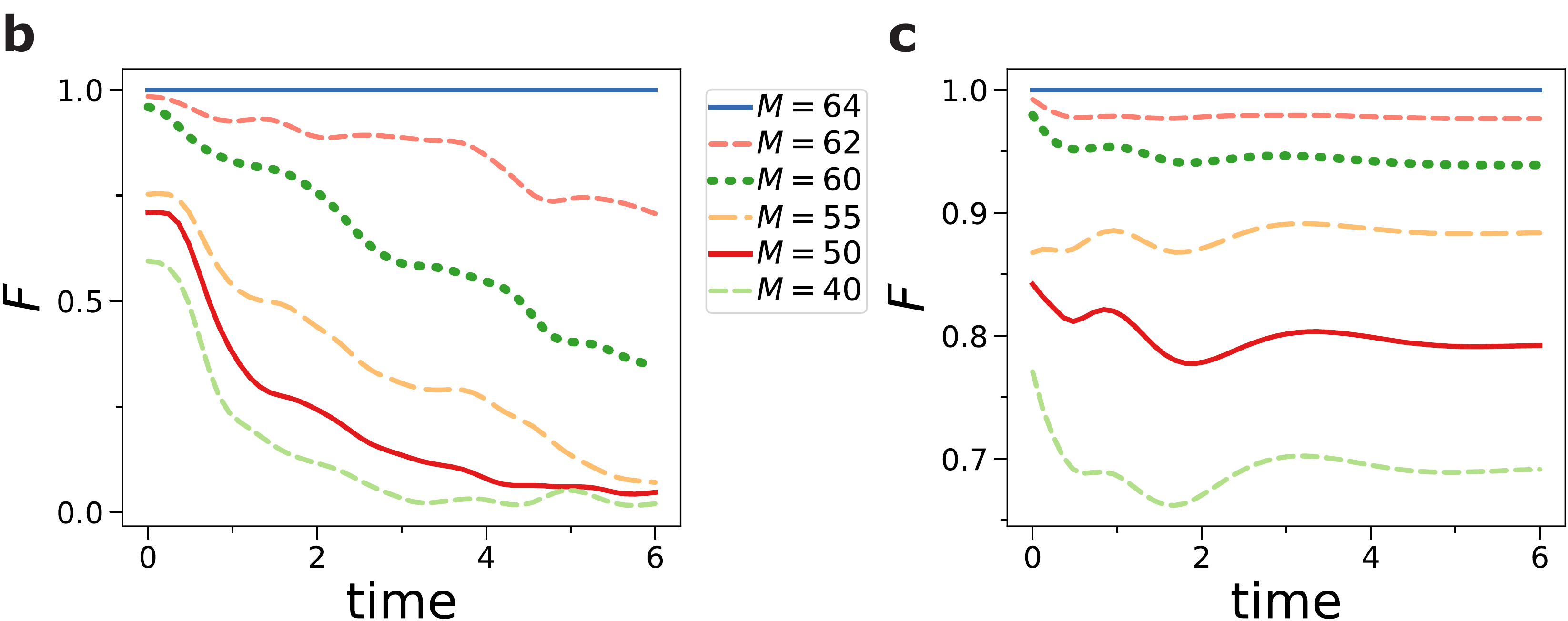}
	\caption{ 
	Simulation of the dynamics of an Ising ladder without and with dissipation. Circuit used to generate moment expansion is shown in Fig.\ref{AnsatzCircuit}. Initial coefficients $\boldsymbol{\beta}(0)$ are calculated via IQAE such that the initial state $\rho(0)$ is a product state with all qubits in state zero.
	Parameters are $N=6$, $J=1$, number of layers $p=6$, $h=1$ and optional dissipative Lindblad term $L_i=\sqrt{\gamma}\sigma_i^+$ with $\gamma=1$.  
	\idg{a} Dynamics of the nearest-neighbor spin correlation $\sum_{\langle i,j \rangle}\sigma_i^z\sigma_j^z/7$ without dissipation (closed) and with disspation (open). Number of basis states is $M=64$.
	\idg{b} Fidelity $F$ of GQAS with exact time evolution for varying number of basis states $M$  without dissipation.
    \idg{c} Fidelity $F$ with dissipation $\gamma=1$.
	}
	\label{ExamleDissp}
\end{figure*}

For the numerical demonstrations in this paper, we generate the reference state $\ket{\psi}$ by a hardware efficient circuit \revC{that is hard to simulate classically}, given by random $y$-rotations followed by control not gates arranged in a hardware-efficient manner in a chain topology (see Fig.\ref{AnsatzCircuit}).

In Fig.\ref{ExamleDissp}a, we plot the evolution of the correlation of nearest-neighbor spins $\sum_{\langle i,j \rangle}\sigma_i^z\sigma_j^z/7$. \revB{The exact dynamics in both closed and dissipative regime can be reproduced exactly for the second $K$-moment expansion. In Fig.\ref{ExamleDissp}b,c we plot the fidelity between the simulation and the exact state as function of time and number of basis states $M$ of the ansatz for closed and open dynamics. We select the $M$ basis states by generating the $2$-moment expansion, and then select the first $M$ states in the order they were generated. \revC{We find similar performance when the states are picked randomly from the $K$-moment expansion.} We find that for $M\ge 64=2^N$, we achieve unit fidelity. For smaller $M$, the ansatz does not cover the full dynamical space, such that the fidelity decreases with time. We find the fidelity improves with increasing $M$. 
We find higher fidelity for open dynamics due to the simulated state being highly mixed, which is easier to represent. The type of reference state is crucial for the representation power of the ansatz and a choice adapted to the problem instead of a randomized circuit could improve the fidelity~\cite{bharti2021nisq}.}

\section{Generalized Time Evolution}\label{sec:gtime}
Next, we want to solve general linear equations with our algorithm. The generalized time evolution is given by
\begin{equation}
B(t)\frac{d}{dt}\vert v(t)\rangle=\vert dv(t)\rangle,\label{eq:GTE_equation}
\end{equation}
such that $\vert dv(t)\rangle=\sum_{j}A_{j}(t)\vert v_{j}^{\prime}(t)\rangle$,
where $A_{j}(t)$ and $B(t)$ are time-dependent
operators and $\vert v(t)\rangle$ is the system state. We use the following ansatz for the system states
\begin{align}
\vert v(t)\rangle=&\sum_{i=0}^{m-1}\alpha_{i}(t)\vert\psi_{i}\rangle\label{eq:System_state_1}\\
\vert v_{j}^{\prime}(t)\rangle=&\sum_{i=0}^{s_{j}-1}\gamma_{j,i}(\boldsymbol{\alpha},t)\vert\phi_{j,i}\rangle\,,\label{eq:System_state_j}
\end{align}
\revB{where we define the state
$\vert v_{j}^{\prime}(t)\rangle$ as a linear combination of arbitrary quantum states $\vert\phi_{j,i}\rangle$ with $s_j$ coefficients $\gamma_{j,i}(\boldsymbol{\alpha},t)$.}
We now use McLachlan's principle for the square of the absolute value and find~\cite{mclachlan1964variational,yuan2019theory}
\begin{multline}
0\stackrel{!}{=}\delta\abs{B(t)\frac{d}{dt}\vert v(t)\rangle-\vert dv(t)\rangle}^2=\\
\sum_{k}\left(\frac{\partial\langle v(t)\vert}{\partial\alpha_{k}^*}B^{\dagger}(t)\left(B(t)\sum_j\frac{\partial\vert v(t)\rangle}{\partial\alpha_{j}}\dot{\alpha}_{j}-\vert dv(t)\rangle\right)\right)\delta\dot{\alpha}_k^*+\\
\sum_{k}\left(\left(\sum_j\frac{\partial\langle v(t)\vert}{\partial\alpha_{j}^*}B^{\dagger}(t)\dot{\alpha}_{j}^*-\langle dv(t)\vert \right)B(t)\frac{\partial\vert v(t)\rangle}{\partial\alpha_{k}}\right)\delta\dot{\alpha}_k
\end{multline}
By demanding that McLachlan's principle is fulfilled for arbitrary variations  $\delta\dot{\alpha}_k$, $\delta\dot{\alpha}_k^*$ and by using $\frac{\partial\vert v(t)\rangle}{\partial\alpha_{j}}=\vert\psi_{j}\rangle$,
we get
\begin{equation}
\sum_{j}\langle\psi_{k}\vert B^{\dagger}(t)B(t)\vert\psi_{j}\rangle\dot{\alpha}_{j}=
\langle\psi_{k}\vert B^{\dagger}(t)\vert v_{j}^{\prime}(t)\rangle.\label{eq:Evolution_equation_2}
\end{equation}
\revB{We assume that $A_{j}(t)$ and $B(t)$ can be written as linear combination of unitaries $U_{j,k}$ and $V_{k}$ with coefficients $\nu_{j,k}(t)$ and $\lambda_{k}$ respectively}
\begin{align}
A_{j}(t)=&\sum_{k}\nu_{j,k}(t)U_{j,k},\label{eq:LCU_Aj}\\
B(t)=&\sum_{k}\lambda_{k}(t)V_{k}.\label{eq:LCU_B}
\end{align}
For the sake of convenience, we define following overlap matrices
\begin{align}
\mathcal{V}_{k,j}\equiv&\langle\psi_{k}\vert B^{\dagger}(t)B(t)\vert\psi_{j}\rangle\label{eq:D_1}\\
\mathcal{D}_{k,j}\equiv&\langle\psi_{k}\vert B^{\dagger}(t)A_{j}(t)\vert v_{j}^{\prime}(t)\rangle.\label{eq:E-1}
\end{align}
We observe that 
\begin{align}
\mathcal{V}_{k,j}=&\sum_{m,n}\lambda_{m}^{\star}(t)\lambda_{n}(t)\langle\psi_{k}\vert V_{m}^{\dagger}V_{n}\vert\psi_{j}\rangle,\label{eq:Overlap_01}\\
\mathcal{D}_{k,j}=&\sum_{j,m,n,p}\lambda_{m}^{\star}(t)\nu_{j,n}(t)\gamma_{j,p}(\boldsymbol{\alpha},t)\langle\psi_{k}\vert V_{m}^{\dagger}U_{j,n}\vert\phi_{j,p}\rangle.\label{eq:Overlap_02}
\end{align}
In terms of the overlap matrices $\mathcal{D}$ and $\mathcal{V}$,
we get the following evolution equation,
\begin{equation}
\mathcal{V}(t)\dot{\boldsymbol{\alpha}}=\mathcal{D}(\boldsymbol{\alpha},t)\, ,\label{eq:Evolution_compact_01}
\end{equation}
where the parameters of $\mathcal{D}(\boldsymbol{\alpha},t)$ can also be function of $\boldsymbol{\alpha}$. 
Real and imaginary time evolution of the Schrödinger equation with a Hamiltonian $H$ is a special case of the GQAS with $B(t)=1$ and $\ket{dv(t)}=\ket{v(t)}$, which has been investigated in \cite{bharti2020quantum2}. For real time evolution, we set $A_1=-iH$ and for imaginary time evolution $A_1=-H$. 

The generalized time evolution can be applied for various other problems, such as linear algebra~\cite{endo2020variational}. Here we show as example how to find the inverse of an invertible matrix $\mathcal{M}$
\begin{equation}
\mathcal{M}\ket{v_{\mathcal{M}^{-1}}}=\ket{v_0}
\end{equation}
with given vector $\ket{v_0}$ and solution to be found $\ket{v_{\mathcal{M}^{-1}}}$. This problem can be converted into a time evolution problem~\cite{endo2020variational} with
\begin{align}
E(t)\ket{v(t)}=&\ket{v_0}\\
E(t)=& \frac{t}{T}\mathcal{M}+\left(1-\frac{t}{T}\right)\mathbb{1}
\end{align}
where $\mathbb{1}$ is the identity matrix, $T$ the final evolution time and $\ket{v(0)}=\ket{v_0}$, $\ket{v(T)}=\ket{v_{\mathcal{M}^{-1}}}$. The derivative of $\ket{v(t)}$ gives the differential equation
\begin{equation}
E(t)\frac{\text{d}}{\text{d}t}\ket{v(t)}=-G(t)\ket{v(t)}
\end{equation}
with $G(t)=(\mathcal{M}-\mathbb{1})/T$. We can identify this equation with Eq.\eqref{eq:GTE_equation}, where $B(t)=E(t)$, $\ket{v'(t)}=\ket{v(t)}$ and $A(t)=-G(t)$.
We assume that $\mathcal{M}=\sum_k\mu_k V_k$ can be represented as a linear combination of unitaries $V_k$. 

We now apply the GQAS algorithm to this problem.
First, we represent the evolving state as a linear combination of states $\ket{v(t)}=\sum_i\alpha_i(t)\ket{\psi_i}$. An efficient way to generate $\ket{\psi_i}$ for a given problem matrix $\mathcal{M}$ could be found using similar methods as the cumulative $K$-moment expansion \cite{bharti2020iterative}.
As second step, the overlap matrices Eqs.\eqref{eq:Overlap_01},\eqref{eq:Overlap_02} are to be measured on the quantum computer. 
For the last step, one integrates Eq.\eqref{eq:Evolution_compact_01} on a classical computer using the measured overlaps for a time $T$ to get $\boldsymbol{\alpha}(T)$ that parameterize the solution vector $\ket{v(t)}=\sum_i\alpha_i(T)\ket{\psi_i}$. 

\section{Nonlinear differential equation}\label{sec:nld}
Nonlinear differential equations are ubiquitous in many areas of science, from hydrodynamic problems such as the Navier-Stokes equations to weather forecasts. It has been recently shown that quantum computers promise exponential speed-up for solving non-linear equations~\cite{lloyd2020quantum}. Further, several other quantum algorithms for non-linear equations have been proposed~\cite{kyriienko2020solving,lubasch2020variational,gaitan2020finding}.
GQAS for generalized time as shown in Eq.\ref{eq:GTE_equation} can be extended to solve non-linear dynamics. Here, we demonstrate the case where $B(t)=1$ with a single non-linear operator $A(t,\vert v(t)\rangle)$. However our algorithm can be easily extended to include linear and non-linear $B(t)$.
We define
\begin{equation}
\frac{d}{dt}\vert v(t)\rangle=A(t,\vert v(t)\rangle)\vert v(t)\rangle,\label{eq:NTE_equation}
\end{equation}
We decompose the operator as linear combination of $r$ unitaries $U_k$ and nonlinear functions $f_k(t,\vert v(t)\rangle)$
\begin{equation}
A(t,\vert v(t)\rangle)=\sum_{k=1}^r f_k(t,\vert v(t)\rangle) U_k
\end{equation}
We now assume that $f_k$ is a nonlinear function of the expectation values of the unitaries $U_k$ with the state $\vert v(t)\rangle$
\begin{equation}
f_k(t,\vert v(t)\rangle)\equiv f_k(t,\langle v(t)\vert U_1\vert v(t)\rangle,\dots,\langle v(t)\vert U_r\vert v(t)\rangle)
\end{equation}

As first step, we define the state $\vert v(t)\rangle$ as a linear combination of basis states $\vert\psi_{i}\rangle$
\begin{equation}
\vert v(t)\rangle=\sum_{i=0}^{m-1}\alpha_{i}(t)\vert\psi_{i}\rangle\label{eq:System_state_NL}\,.
\end{equation}
As second step, one uses a quantum computer to measure the overlaps $\mathcal{E}$ as defined in Eq.\eqref{eq:open_E} and
\begin{align}
\mathcal{S}_{i,j}^k=\langle\psi_{i}\vert U_k\vert\psi_{j}\rangle\, . \label{eq:open_S}
\end{align}
As third and final step, we rewrite Eq.\eqref{eq:NTE_equation} in terms of above definitions to find
\begin{equation}
\mathcal{E}\frac{\text{d}}{\text{d}t}\boldsymbol{\alpha}=-i\left(\sum_k f_k(t,\boldsymbol{\alpha}^\dagger\mathcal{S}^1\boldsymbol{\alpha},\dots,\boldsymbol{\alpha}^\dagger\mathcal{S}^r\boldsymbol{\alpha})\mathcal{S}^k\right)\boldsymbol{\alpha}\label{eq:NLQAS}\,.
\end{equation}
This nonlinear differential equation is then solved on a classical computer.

\begin{figure}[htbp]
	\centering
	\subfigimg[width=0.45\textwidth]{}{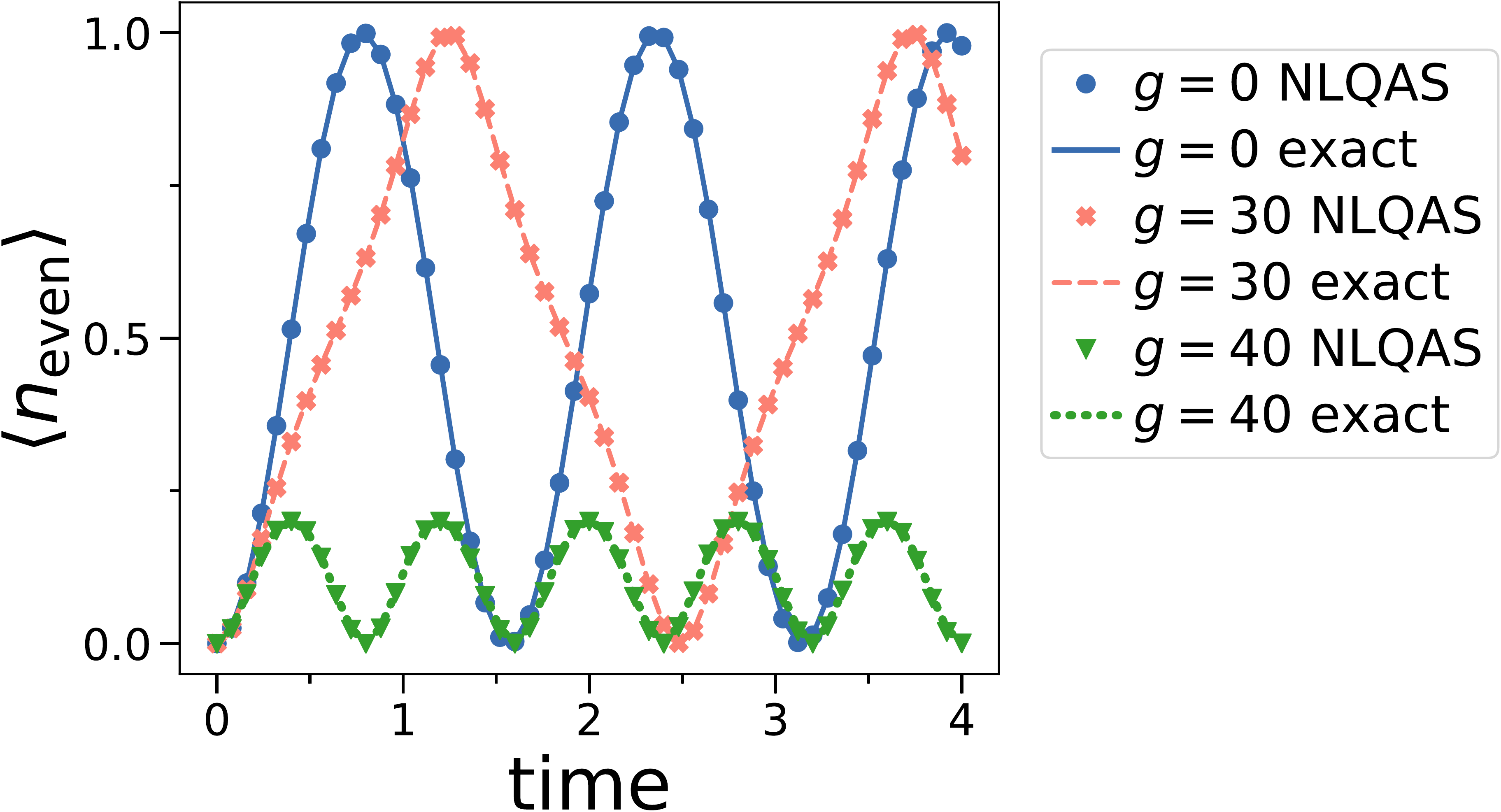}
\caption{Evolution of the density for even number of sites $\langle n_\text{even}\rangle$ for varying nonlinear parameter $g$ of the discrete nonlinear Schrödinger equation (Eq.\eqref{eq:NLschrodinger}). 
Parameters of the system are $N=8$, $V_i=0$ and $J=1$.}
	\label{ExampleNL}
\end{figure}

We now demonstrate how to solve the non-linear Schrödinger equation~\cite{scott2003nonlinear} with our method. This type of equation is for example used in nonlinear optics and describes Bose-Einstein condensates with weakly interacting particles. 
A simple discrete version of this equation in one dimension is given by
\begin{equation}
i\frac{\text{d}}{\text{d}t}\eta_i=-J(\eta_{i+1}+\eta_{i-1})+V_i\eta_{i}+g\abs{\eta_{i}}^2\eta_{i}\, ,\label{eq:NLschrodinger}
\end{equation}
where $\eta_i$ are $N$ complex numbers  that are normalized with $\sum_i\abs{\eta_i}^2=1$, $J$ is the coupling strength of neighboring discrete states $i$ and $i+1$, $V_i$ the local energy and $g$ the non-linear interaction strength. 
We encode the $\eta_i$ into a quantum state via $\ket{\phi}=\sum_i\eta_i\sigma^x_i\ket{0}$. Now, the discrete nonlinear Schrödinger equation can be mapped to a spin Hamiltonian of $N$ qubits with nonlinear coefficients, under the condition that the state $\ket{\phi(t)}$ is an eigenstate of $M=\sum_i n_i$ with $M\ket{\phi}=\ket{\phi}$ with eigenvalue $\lambda_M=1$, where we define the density $n_i=\frac{1}{2}(\mathbb{1}-\sigma^z_i)$.
We now define the nonlinear spin Hamiltonian 
\begin{equation}
H=-\frac{J}{2}\sum_i(\sigma^x_i\sigma^x_{i+1}+\sigma^y_i\sigma^y_{i+1})+
\sum_i V_in_i+g\sum_i \bra{\phi}n_i\ket{\phi}n_i\,.\label{eq:SchrodingerSpin}
\end{equation}
We can now solve the discrete non-linear Schrödinger equation (Eq.\ref{eq:NLschrodinger}) by mapping it to Eq.\eqref{eq:SchrodingerSpin}. 
In Fig.\ref{ExampleNL}, we simulate both the linear regime $(g=0)$ as well as the nonlinear dynamics $(g>0)$. We use an ansatz a linear combination of the states $\ket{\psi_i}=\sigma_i^x\ket{0}$. The initial state of the simulated evolution at $t=0$ is a superposition state with non-zero density at odd number of sites, e.g. $\sum_{i\in \text{odd}}\langle n_i\rangle=1$ and zero density at even sites $\langle n_\text{even}\rangle=\sum_{i\in \text{even}}\langle n_i\rangle=0$, with density at site $i$ $n_i=\frac{1}{2}(-\sigma_i^z+1)$.
We observe with increasing $g$ that the density oscillation is suppressed, demonstrating the onset of non-linear behavior and self-trapping~\cite{albiez2005direct}.

\section{Quantum Assisted Gibbs State Preparation}\label{sec:gibbs}
Preparation of quantum states at a given temperature is an important task relevant for many quantum algorithms. However, generating these Gibbs state can be quite challenging. A common approach is to evolve the totally mixed state in imaginary time~\cite{verstraete2004matrix}. However, in general preparing the totally mixed state is very resource demanding. Further, imaginary time evolution requires non-unitary dynamics, which can be hard to implement on quantum computers.
Here, we propose to approximate the totally mixed state using the hybrid density matrix and apply QAS as primitive to implement the imaginary time evolution to generate Gibbs states.

The Gibbs state is given by 
\begin{equation}\label{eq:Gibbs}
\rho(T)=\frac{e^{-H/T}}{\Tr(e^{-H/T})}\,,
\end{equation}
where $H$ is some Hamiltonian and $T$ the temperature.
To get this state, the totally mixed state $\rho_I=\mathbb{1}_\mathcal{N}/\mathcal{N}$ can be evolved in imaginary time $\tau$ to get the state with temperature $T=1/(2\tau)$. 
The evolution of a density matrix in imaginary time is given by
\begin{equation}\label{eq:OpenImag}
\frac{\text{d}}{\text{d}\tau}\rho=-(H\rho+\rho H)\, .
\end{equation}
We now solve this equation using QAS. 
First, we expand $\rho=\sum_{i,j}\beta_{i,j}(t)\ket{\psi_i}\bra{\psi_j}$ as hybrid density matrix corresponding to  Eq.\eqref{eq:open_ansatz} with combination parameters $\beta$. The basis states $\ket{\psi_i}$ can be for example generated via $K$-moment expansion (see Eq.\eqref{eq:kmoment}). 
Then, we measure the overlap matrices $\mathcal{E}$ and $\mathcal{D}$ as defined in Eq.\eqref{eq:open_E},\eqref{eq:open_D} on a quantum computer.
Finally, we solve the corresponding equations for imaginary time evolution 
\begin{equation}
\mathcal{E}\frac{\text{d}}{\text{d}\tau}\boldsymbol{\beta}\mathcal{E}=-(\mathcal{D}\boldsymbol{\beta}\mathcal{E}+\mathcal{E}\boldsymbol{\beta}\mathcal{D})\label{eq:gibbs_E}\, .
\end{equation}
For the initial parameters for the hybrid density matrix, we choose $\beta(\tau=0)=\beta_\text{I}$ to approximate the totally mixed state with
\begin{equation}
\beta_\text{I}=\frac{\mathcal{E}^+}{\Tr(\mathcal{E}^+\mathcal{E})}\, ,\label{eq:totallymixed}
\end{equation}
where $\mathcal{E}^+$ is the pseudo inverse of $\mathcal{E}$. 

\begin{figure}[htbp]
	\centering
	\subfigimg[width=0.35\textwidth]{}{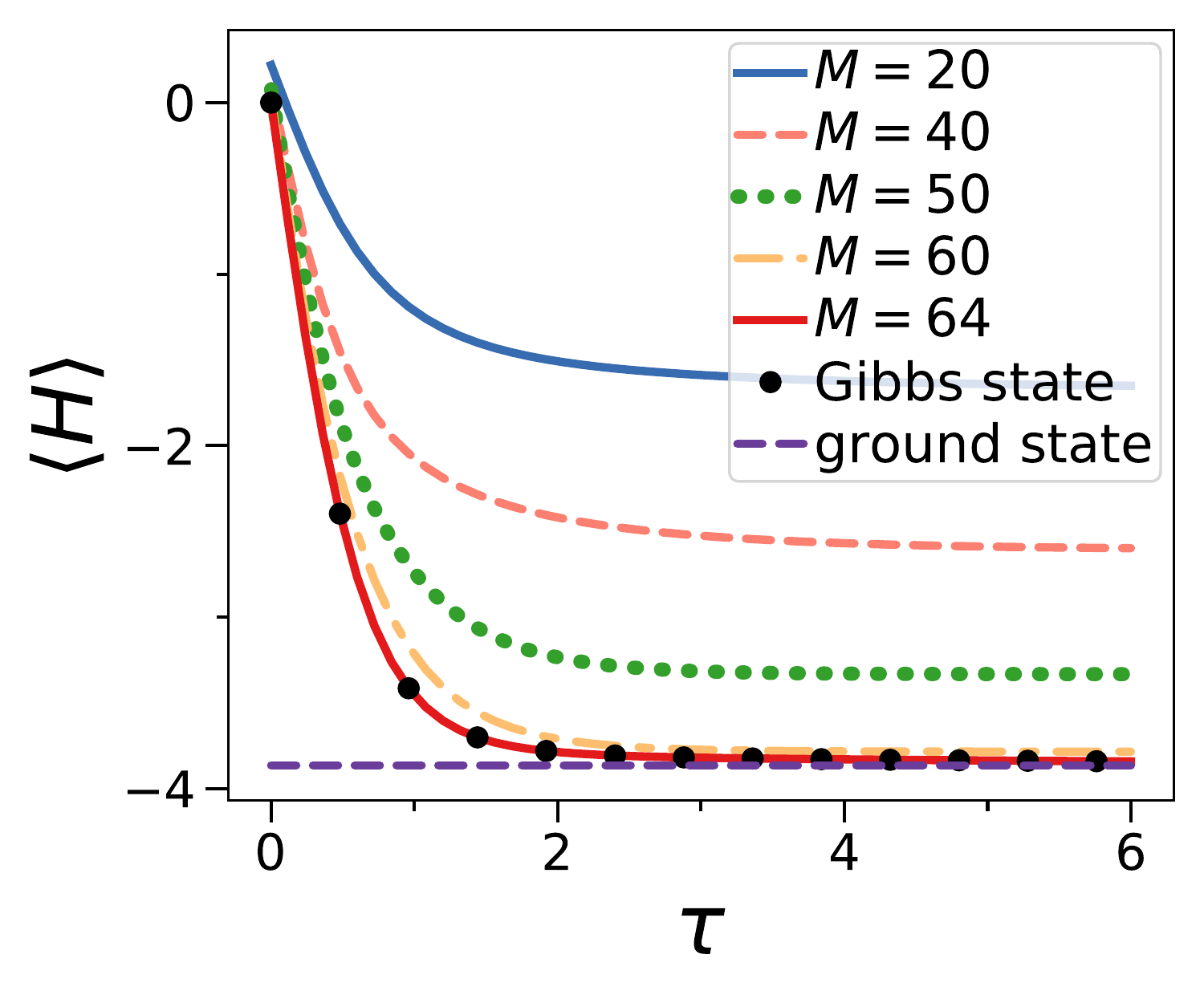}
	\caption{Preparation of Gibbs state $\rho(T)=\exp(-H/T)$ with temperature $T$ for Ising model Hamiltonian $H$ (Eq.\eqref{eq:ising}) by evolving the totally mixed state $\rho_\text{I}=\mathbb{1}_\mathcal{N}/\mathcal{N}$ in imaginary time $\tau=T/2$. We show the expectation value of energy $\langle H(\tau)\rangle$ for varying number of basis states $M$. 
	The Hamiltonian is the transverse Ising model given by Eq.\eqref{eq:ising} with parameters $N=6$, $p=6$, $J=1$ and $h=1$.}
	\label{ExampleGibbs}
\end{figure}

As example, we show in Fig.\ref{ExampleGibbs} the preparation of the Gibbs state with QAS for different moment expansion $K$ of the transverse field Ising model
\begin{equation}\label{eq:ising}
H_\text{ising}=\frac{J}{2}\sum_{i=1}^N\sigma^x_i\sigma^x_{i+1}-\frac{h}{2}\sum_{i=1}^N\sigma^z_i
\end{equation}
with nearest-neighbor coupling $J$ and transverse field $h$. 
We approximate the totally mixed state as hybrid density matrix $\rho=\sum_{i,j}\beta_{i,j}(t)\ket{\psi_i}\bra{\psi_j}$ with coefficients given by Eq.\eqref{eq:totallymixed}. The states $\ket{\psi_i}$ are generated from the $K$-moment expansion (see Eq.\eqref{eq:kmoment}) using a hardware efficient circuit as basis of the expansion as shown in Fig.\ref{AnsatzCircuit}. We select $M$ basis states in the order they created in the $K$-moment expansion.
The hybrid density matrix is evolved in imaginary time using Eq.\eqref{eq:gibbs_E}. As reference, we also show ground state energy, to which the Gibbs state converges in the limit of $\tau\rightarrow\infty$.
The simulation converges to the exact Gibbs state with increasing $M$, reaching the exact state for $M=64$ and allowing us to prepare Gibbs state with arbitrary temperature $T$ by varying the evolution time $\tau$.

Alternatively, one can start with maximally entangled state $\ket{\zeta}_{AB}$ of system AB, evolve the whole system under the Hamiltonian $H_A \otimes I_B$ via imaginary time evolution using QAS for time $\tau$~\cite{yuan2019theory}. The state of system A at time $\tau$ is then given by the Gibbs state with temperature $T=\frac{1}{2\tau}$.

\section{Discussion and Conclusion}\label{sec:discussion}
In this work, we presented a NISQ era algorithm to simulate open system dynamics, generalized time evolution, nonlinear differential equations and Gibbs state preparation. The GQAS proceeds in three steps. 
The first step selects the ansatz as a linear combination of basis states $\ket{\psi_i}\in\mathbb{S}$ from a set $\mathbb{S}$. The ansatz can be either a pure state or mixed state $\rho$, depending on the problem. Here, we introduce the concept of hybrid density matrix (Eq.\eqref{eq:open_ansatz}) to represent a mixed state as a linear combination of parameters, where the combination parameters can be updated on a classical computer without the need of tuning the quantum states. 
The second step involves the computation of overlap matrices on a quantum computer, which can be performed efficiently using techniques of \cite{mitarai2019methodology} or by assuming construction from Pauli strings~\cite{bharti2020quantum2}. After the overlap matrices have been computed, the differential equation to be computed is solved on a classical computer. 
Note that once the ansatz has been decided, our algorithm does not mandate any quantum-classical feedback loop as the quantum states defining the ansatz are fixed, and only the variational parameters are (classically) updated. The algorithm does not require the computation of gradients using the quantum computer and thus circumvents the barren plateau problem by construction.

\revC{Our algorithm simulates the dynamics within an ansatz space spanned by $M$ states. The ansatz states are prepared on a quantum computer and we measure corresponding overlaps. By choosing ansatz states that are intractable for classical simulation, our algorithm has the potential to exceed classical simulation methods. 
Our algorithm achieves high fidelity when the ansatz spans the subspace of the dynamics (see Fig.\ref{ExamleDissp}b,c), else we observe a decay in fidelity with increasing time. 
To run our algorithm with high fidelity, a main challenge is to keep the number of ansatz states small by choosing the most important states. In the worst case, an exponential number of states is needed when the dynamics is highly ergodic and spans the whole Hilbertspace. For short times or constrained problems with many-body scars~\cite{serbyn2021quantum}, symmetries or many-body localization, dynamics is restricted to a small part of the Hilbertspace, which can be spanned by a polynomial number of states.  For simulating dynamics, we use the cumulative $K$-moment expansion with Eq.\eqref{eq:kmoment} to construct the ansatz. This method is inspired by the Krylov subspace expansion, a powerful classical method for simulation that uses the Hamiltonian to find a suitable subspace~\cite{saad1992analysis}. 
We highlight that the cumulative $K$-moment expansion has been recently shown to find a good subspace  with polynomial scaling for a spin model by choosing an reference state inspired by quantum annealing~\cite{bharti2021nisq}. As another example, for fermionic problems one can use fermionic excitation operators to build an ansatz space that covers the main excitations of the problem~\cite{ollitrault2020quantum}. We believe by combining an initial state as well as expansion operators tailored to the problem, a sub-exponential scaling could be found for many problems.

For the classical post-processing step, our algorithm requires inversion of the $\mathcal{E}$ matrix. If $\mathcal{E}$ has small eigenvalues, the inversion can be sensitive to experimental noise, which can negatively impact the accuracy. Recent work has found improved methods to reduce the error of the inversion  step~\cite{epperly2021theory}. }

\revB{Note that the overlaps corresponding to equations~\eqref{eq:open_D},~\eqref{eq:open_R} and~\eqref{eq:open_F} can be exponentially small when using random circuits to generate $\vert \psi_i \rangle$. However, we would like to emphasise that this problem corresponds to the calculation of expectation values for random quantum states and is fundamentally different from that of gradients. For random circuits, one can have both exponentially small expectation values as well as vanishing gradients. The former problem exists in our approach as well as variational quantum algorithms. However, our approach bypasses the vanishing gradient problem by construction as no gradients are measured with the quantum computer.}
The measurements involved in the estimation of the overlap matrices can be performed efficiently without the requirement of any complicated measurements, such as the Hadamard test. Refer to the Appendix of reference \cite{bharti2020iterative} for details.
The GQAS algorithm can trivially subsume the algorithms based on VQE (or VQS) by allowing the quantum states defining the ansatz to be variationally adjusted. See the Appendix for an illustration. Apart from straightforward applications, our algorithm could be used as primitive for more involved algorithms such as quantum Boltzmann training, quantum signal processing and algorithms for quantum machine learning. Our algorithm also harbour potential applications in solving the Navier-Stokes equation and plasma hydrodynamics.

The algorithms designed for fault-tolerant quantum computers, such as Shor's factoring and Grover search, allow a rigorous mathematical analysis.  On the contrary, algorithms such as VQE and VQS, which do not require fault-tolerant architecture, are heuristic in nature and often render any systematic analysis challenging.  A borderline exception is the quantum approximate optimization algorithm~\cite{farhi2014quantum}, which can be analyzed though its implementation can be challenging on the existing quantum hardware. 
In contrast, our algorithm are compatible with the existing NISQ capabilities and provide proper
mathematical analysis of its optimization program.

Most of the quantum computers nowadays are being accessed via cloud platforms by submitting jobs into a queue, which is then executed at some later time. The classical-quantum feedback loops of VQE require a lot of time, as each step of the loop has to wait for the previous job to finish in the queue. Only then, one can perform the classical part of the algorithm and submit the next job for the quantum computer. The sequential nature of classical-quantum feedback loop severely slows down the execution these algorithms.
In contrast, our algorithm do not require any classical-quantum feedback loop and are embarrassingly parallel as all quantum computations are independent of each other, allowing for much faster termination.

In future, it would be interesting to study GQAS algorithms in the presence of noise. A proper error analysis is in order. Providing complexity-theoretic guarantees would be another exciting direction.

\medskip
{\noindent {\em Acknowledgements---}} We are grateful to the
National Research Foundation and the Ministry of Education, Singapore for financial support.

\bibliography{GQAS}

\appendix

\newpage
\section{Quantum Assisted Variational Simulator} \label{sec: QAVS}
One can select a linear combination of variational quantum states
and thus allow to update the quantum states defining the ansatz. The
aforementioned strategy furnishes a method to allow QAS based algorithms
to subsume VQS based algorithms. For the sake of illustration, let
us consider the ansatz state as 
\[
\vert\phi\left(\boldsymbol{\theta},\boldsymbol{\alpha}\right)\rangle=\sum_{i=1}^{M}\alpha_{i}\vert\psi_{i}\left(\boldsymbol{\theta}\right)\rangle,
\]
where $\vert\psi_{i}\left(\boldsymbol{\theta}\right)\rangle=U_{i}\vert\psi\left(\boldsymbol{\theta}\right)\rangle$
for some unitary $U_{i}$ (constructed using the unitaries defining
the underlying Hamiltonian) and $\boldsymbol{\theta}=\theta_{1}\cdots\theta_{K}.$
We assume the variational parameters to be real valued and leave the
more detailed analysis for involved cases such as $\alpha_{i}\in\mathbb{C}$
and $\theta_{i}\in\mathbb{R}$ for future works. Simple algebraic
calculations based on McLachlan principle yields the following update
equation,

\begin{equation}
\left[\begin{array}{cc}
P & Q\\
S & T
\end{array}\right]\left[\begin{array}{c}
\dot{\boldsymbol{\theta}}\\
\dot{\boldsymbol{\alpha}}
\end{array}\right]=\left[\begin{array}{c}
R\\
W
\end{array}\right].\label{eq:Update_subsume}
\end{equation}
Here, $P,Q,S,T,R$ and $W$ are $K\times K,K\times M,M\times K,M\times M,K\times1$
and $M\times1$ matrices with the following description,

\[
P_{i,j}=\frac{\partial\langle\phi\vert}{\partial\theta_{i}}\frac{\partial\vert\phi\rangle}{\partial\theta_{j}}+\frac{\partial\langle\phi\vert}{\partial\theta_{j}}\frac{\partial\vert\phi\rangle}{\partial\theta_{i}},
\]

\[
Q_{i,j}=\frac{\partial\langle\phi\vert}{\partial\theta_{i}}\vert\psi_{j}\rangle+\langle\psi_{j}\vert\frac{\partial\vert\phi\rangle}{\partial\theta_{i}},
\]

\[
S_{i,j}=\langle\psi_{i}\vert\frac{\partial\vert\phi\rangle}{\partial\theta_{j}}+\frac{\partial\langle\phi\vert}{\partial\theta_{j}}\vert\psi_{i}\rangle,
\]

\[
T_{i,j}=\langle\psi_{i}\vert\psi_{j}\rangle+\langle\psi_{j}\vert\psi_{i}\rangle,
\]

\[
R_{i}=-\iota\left(\frac{\partial\langle\phi\vert}{\partial\theta_{i}}H\vert\phi\rangle-\langle\phi\vert H\vert\frac{\partial\vert\phi\rangle}{\partial\theta_{i}}\right),
\]
and

\[
W_{i}=-\iota\left(\langle\psi_{i}\vert H\vert\phi\rangle-\langle\phi\vert H\vert\psi_{i}\rangle\right).
\]
Here, for the sake of brevity, we have denoted $\vert\phi\left(\boldsymbol{\theta},\boldsymbol{\alpha}\right)\rangle$
by $\vert\phi\rangle,$ and $\vert\psi_{i}\left(\boldsymbol{\theta}\right)\rangle$
by $\vert\psi_{i}\rangle.$
\end{document}